\documentclass[aps,preprint]{revtex4}
\usepackage{amsmath,amsfonts,amssymb,amsthm}

\newtheorem{Lemma}{Lemma}
\newtheorem{Theorem}{THEOREM}
\newtheorem{Proposition}{PROPOSITION}
\newtheorem{corollary}{COROLLARY}

\newcommand{\R}{\mathbb{R}}
\renewcommand{\S}{\mathbb{S}}

\newcommand{\F}{\mathcal{F}}
\newcommand{\W}{\mathcal{W}}

\newcommand{\gm}{\gamma}
\newcommand{\al}{\alpha}

\newcommand{\half}{\mbox{$\frac 12$}}

\newcommand{\be}{\begin{equation}}
\newcommand{\ee}{\end{equation}}
\newcommand{\bea}{\begin{align}}
\newcommand{\eea}{\end{align}}

\newcommand\infspec{{\rm{inf\, spec\,}}}
\newcommand\eps\epsilon
\newcommand\V{\mathcal{V}}
\newcommand\B{\mathcal{B}}
\newcommand\dig{\mathfrak{F}}

\begin{document}
\title{Critical Temperature and Energy Gap for the BCS Equation}

\author{Christian Hainzl} 
\affiliation{Departments of
  Mathematics and Physics, UAB, 1300 University Blvd, Birmingham AL
  35294, USA} \email{hainzl@math.uab.edu}

\author{Robert Seiringer} 
\affiliation{Department of
  Physics, Princeton University, Princeton NJ 08542-0708, USA}
\email{rseiring@princeton.edu}

\date{May 23, 2008}

\begin{abstract}
  We derive upper and lower bounds on the critical temperature
  $T_c$ and the energy gap $\Xi$ (at zero temperature) for the BCS gap
  equation, describing spin $1/2$ fermions interacting via a local
  two-body interaction potential $\lambda V(x)$.  At weak coupling
  $\lambda \ll 1$ and under appropriate assumptions on $V(x)$, our bounds
  show that $T_c \sim A \exp(-B/\lambda)$ and $\Xi \sim C
  \exp(-B/\lambda)$ for some explicit coefficients $A$, $B$ and $C$
  depending on the interaction $V(x)$ and the chemical potential
  $\mu$. The ratio $A/C$ turns out to be a universal constant,
  independent of both $V(x)$ and $\mu$. Our analysis is valid for any
  $\mu$; for small $\mu$, or low density, our formulas reduce to
  well-known expressions involving the scattering length of $V(x)$.
\end{abstract}


\maketitle

\section{Introduction}

The recent advances in trapping and cooling of cold atoms have led to
renewed interest in the behavior of ultra-cold fermionic gases. Under
the assumption that the interactions among the individual fermions are
weak, the system shows a superfluid behavior at low temperature that 
is well described \cite{Leg,NS,rand,andre,PMTBL,Chen,zwerger,CCPS} by the
Bardeen-Cooper-Schrieffer (BCS) model \cite{BCS}. The BCS model was
originally introduced as a model for electrons displaying
superconductivity, and has played a prominent role in condensed matter
physics in the fifty years since its introduction.  We shall not be
concerned here with a mathematical justification of the approximations
leading to the BCS model, but rather with an investigation of its
precise predictions.

In this paper, we study the BCS gap equation for fermionic systems
with general local interaction potentials. We give a rigorous
derivation of expressions for the {\it critical temperature} $T_c$ and
the {\it energy gap} $\Xi$ (at zero temperature) that are valid to
second order Born approximation. More precisely, for all interaction
potentials $V(x)$ that create a negative energy bound state of the
effective potential on the Fermi sphere (see Eq. (\ref{defvm}) below;
a sufficient condition for this property is that $\int_{\R^3} V(x) dx
< 0$), we show that
\begin{equation}\label{form:tc}
T_c = \mu \frac{8 e^{\gamma-2}}{\pi} e^{\pi/(2 \sqrt{\mu} b_\mu)} 
\end{equation}
where $\mu>0$ is the chemical potential, $\gamma\approx 0.577$ denotes
Euler's constant, and $b_\mu<0$ is an effective scattering
length. Units are chosen such that $\hbar=k_{\rm B}=2m = 1$, where $m$ is the
mass of the fermions. To first order in the Born approximation,
$b_\mu$ is related to the scattering amplitude of particles with
momenta on the Fermi sphere, but to second order the expression is
more complicated. The precise formula is given in Eq.~(\ref{bmexpl})
below.  For interaction potentials that decay fast enough at large
distances, we shall show that $b_\mu$ reduces to the {\it scattering length} $a_0$
of the interaction potential in the low density limit, i.e., for small
$\mu$.

We emphasize that not only are our results mathematically rigorous,
but our analysis holds for arbitrary (positive) values of the chemical
potential $\mu$. In particular, the formula (\ref{form:tc})
generalizes previously known results valid only for small $\mu$, i.e.,
low density. To the best of our knowledge, the correct expression for
$b_\mu$ does not seem to have appeared in the literature before. We
use precise spectral analysis of the (linearized) BCS gap equation in
order to derive our results. 

For simplicity, we restrict our analysis to local interaction
potentials, which is the case of interest when describing dilute Fermi
gases. Our methods are applicable in a much more general setting,
however, and generalizations to non-local potentials (as used in the
theory of superconductivity) are straightforward.

For interaction potentials that have non-positive Fourier transform
(ensuring, in particular, that the BCS pair wavefunction is unique and
has zero angular momentum), we shall prove similar results for the
zero temperature energy gap, which we denote by $\Xi = \min_p
\sqrt{(p^2-\mu)^2 + |\Delta(p)|^2}$. It turns out that, at least up to second
order Born approximation,
\begin{equation}\label{form:xi}
\Xi = T_c \frac \pi{e^\gamma}
\end{equation}
in this case. This equality is valid for any density, i.e., for any
value of the chemical potential $\mu$. In particular, $\Xi$ has exactly the
same exponential dependence on the interaction potential, described by
$b_\mu$, as the critical temperature $T_c$.

Despite the huge physics literature concerning the
BCS gap equation, rigorous results concerning its prediction are
sparse. Our analysis here relies on the previous studies in
\cite{HHSS} in this direction, and extends the results of \cite{FHNS}.

Before giving the mathematically precise statements in Section~\ref{sec:precise},
we shall give a non-technical discussion of our main results in the
next section. Sections~\ref{proof:thm1}--\ref{sec:gap} contain their proofs.

\section{Discussion of Main Results}

We consider a gas of spin $1/2$ fermions at temperature $T\geq 0$ and
chemical potential $\mu >0$, interacting via a local two-body
interaction potential of the form $2\lambda V(x)$. Here, $\lambda>0$
is a coupling parameter, and the factor $2$ is introduced for
convenience. The superfluid phase of the system is described by the BCS gap equation
\begin{equation}\label{bcsex}
  \Delta(p) = -\frac \lambda{(2\pi)^{3/2}} \int_{\R^3} \hat V(p-q)
  \frac{\Delta(q)}{E(q)} \tanh\frac{E(q)}{2T} \, dq\,,
\end{equation}
where $E(q)=\sqrt{(p^2-\mu)^2+|\Delta(q)|^2}$.  The Fourier transform
of $V(x)$ is denoted by $\hat V(p)$.

We are interested in the critical temperature $T_c$ which, in
mathematical terms, is defined by the property that (\ref{bcsex}) has
a non-trivial (that is, not identically vanishing) solution for $T<
T_c$, while there is no solution for $T\geq T_c$. In the limit
$\lambda\to 0$, we shall show in Theorem~\ref{constant} that
  \begin{equation}\label{themeq2x}
    \lim_{\lambda \to 0} \left(\ln\left(\frac\mu {T_c}\right) + 
      \frac {\pi}{2 \sqrt{\mu}\, b_\mu(\lambda)}\right) = 2 - \gamma - \ln(8/\pi)\,.
  \end{equation}
  Here, $\gamma\approx 0.577$ denotes Euler's constant. For small
  $\lambda$, $T_c$ is thus given by (\ref{form:tc}). The quantity
  $b_\mu(\lambda)$ plays the role of an effective scattering
  length. In fact, as $\mu\to 0$, it reduces exactly to the scattering
  length of $2\lambda V(x)$. For general $\mu$, the expression is slightly more
  complicated, however. The precise formula is given in the next
  section, see Eq.~(\ref{bmexpl}). In certain special cases, including
  all potentials with non-positive Fourier transform, $b_\mu(\lambda)$
  is given by
\begin{align}\nonumber
  b_\mu(\lambda) &= \frac{\lambda}{4\pi} \int_{\R^3} V(x)
  \frac{\sin^2(\sqrt\mu |x|)}{\mu|x|^2} dx \\ &\quad +\frac{\pi
    \lambda^2}{2} \int_{\R^3} \left( \frac { |\hat\varphi(p)|^2 -
      |\hat \varphi(\sqrt{\mu})|^2}{|p^2-\mu|} + \frac{1}{p^2}
    |\hat\varphi(\sqrt\mu)|^2\right) dp +O(\lambda^3) \label{bmu}
\end{align}
where $\varphi(x)=(2\pi^2\mu)^{-1/2} V(x) \sin(\sqrt\mu|x|)/|x|$ and
$\hat\varphi(p)$ is its Fourier transform. 

For $\Delta$ the solution of the gap equation (\ref{bcsex}) at zero
temperature $T=0$, the energy gap of the system is given by $\Xi
=\min_{p} \sqrt{(p^2-\mu)^2 + |\Delta(p)|^2}$. For small $\lambda$, it
shows the same behavior as the critical temperature. More precisely,
we shall show in Theorem~\ref{gap} that
\begin{equation}\label{themeq2y}
  \lim_{\lambda \to 0} \left(\ln\left(\frac\mu \Xi \right) + 
    \frac {\pi}{2 \sqrt{\mu}\, b_\mu(\lambda)}\right) = 2 - \ln(8)\,.
\end{equation}
Together with (\ref{themeq2x}) this shows the validity of (\ref{form:xi}) for small $\lambda$. 

We emphasize that our results hold for a large class of interaction
potentials $V(x)$ and are hence suitable to describe a wide range of
physical situations. It is not necessary to make the approximation of
zero-range, for instance, as is often done in the physics
literature. Our results can therefore be interpreted as an {\it a
  posteriori} justification of such an approximation. Note that $V(x)$
does not necessarily have to be interpreted as the exact interaction
potential among the particles but can include effective interactions
arising from higher order contributions not taken into account in the
BCS approximation.

Formulas (\ref{form:tc}) and (\ref{form:xi}) are well-known for small $\mu$, with
$b_\mu(\lambda)$ replaced by the scattering length of $2\lambda V(x)$,
which we shall denote by $a_0(\lambda)$. Our analysis shows that they
are valid for all $\mu>0$.  For given potential $V(x)$, the effect of non-zero $\mu$ on
$b_\mu(\lambda)$ can easily be calculated. For instance, for a
Gaussian potential $V(x)= - \kappa \exp(-\kappa |x|^2)$, the difference
between $b_\mu(\lambda)$ and the scattering length $a_0(\lambda)$ of
$2\lambda V$ is given by
$$
b_\mu(\lambda)- a_0(\lambda) =- \frac \mu {\kappa^{3/2}} \left(
  \frac{\sqrt\pi}4 \lambda + {\sqrt\pi}\left[ \ln(1+\sqrt 2)
    +\frac 1{2 \sqrt2}\right] \lambda^2 + O(\lambda^3) \right) \
\text{ for $\mu \ll \kappa$}\,.
$$ 
Higher order corrections can  be calculated from (\ref{bmu}) as well.

\section{Precise Statement of Results}\label{sec:precise}

In the following, we shall assume that $V$ is a real-valued potential
that has some mild regularity properties, namely $V\in L^1(\R^3)\cap
L^{3/2}(\R^3)$, i.e., $\int_{\R^3} |V(x)|^p dx < \infty$ for $1\leq p
\leq 3/2$. In the BCS approximation, the system under consideration is
described by the {\it BCS functional} $\F$, which was introduced by
Leggett in \cite{Leg}. If $\gamma$ denotes the momentum distribution
of the fermions, and $\alpha$ is the pair wavefunction, $\F$ is given
by \cite{HHSS}
\begin{equation}\label{bcsfunc}
  \F(\gamma,\alpha) = \int_{\R^3} (p^2-\mu)\gm(p)dp+ \lambda \int_{\R^3} |\alpha(x)|^2
  V(x)dx - T S(\gm,\al)\,.
\end{equation}
Here, $S$ denotes the entropy functional 
$$
S(\gm,\al)=-\int_{\R^3} \left[
  s(p)\ln s(p)+\big(1-s(p)\big)\ln\big(1-s(p)\big)\right] dp\,,
$$ 
with $s(p)$ determined by $s(1-s)=\gm(1-\gm)-|\hat\al|^2$. Here and in
the following, we use a caret to denote Fourier transform; i.e.,
$\hat f(p) = (2\pi)^{-3/2} \int_{\R^3} f(x) e^{-ipx} dx$. The BCS
functional $\F$ is naturally defined for pairs of functions
$(\gamma,\alpha)$ with $\gamma\in L^1(\R^3,(1+p^2)dp)$, $0\leq
\gamma(p)\leq 1$, and $\alpha\in H^1(\R^3,dx)$ satisfying
$|\hat\alpha(p)|^2\leq \gamma(p)(1-\gamma(p))$.

The existence of a minimizer of $\F$ was shown in \cite{HHSS}. The
corresponding variational equation satisfied by a minimizer can be
formulated as follows. Given a minimizing pair $(\gamma,\alpha)$, one
defines $\Delta$ as
\begin{equation}\label{defdelta}
  \Delta(p) = \frac{p^2-\mu}{\half-\gamma(p)}\hat\alpha(p)\,.
\end{equation}
It satisfies the {\it BCS gap equation}
\begin{equation}\label{bcse}
  \Delta(p) = -\frac \lambda{(2\pi)^{3/2}} \int_{\R^3} \hat V(p-q)
  \frac{\Delta(q)}{E(q)} \tanh\frac{E(q)}{2T} \, dq
\end{equation}
with $E(p)= \sqrt{(p^2-\mu)^2 + |\Delta(p)|^2}$. If $\Delta$ does not
vanish identically (or, equivalently, the minimizing $\alpha$ does not
vanish identically), the system is said to be in a superfluid phase.

\subsection{Critical Temperature}\label{subsec:ct}

It was shown in \cite{HHSS} that there exists a {\it critical
  temperature} $T_c$ below which the gap equation (\ref{bcse}) has a
non-trivial solution (i.e., $\Delta$ does not vanish identically), and
above which it does not. Equivalently, the $\alpha$ minimizing the BCS
functional $\F$ is identically zero for $T\geq T_c$, while it is
non-zero for $T<T_c$. This critical temperature is characterized by
\cite[Thms.~1 and 2]{HHSS}
\begin{equation}\label{deftc}
  T_c = \inf \left\{ T>0\, : \, \infspec \left( K_{T,\mu} + \lambda V \right) \geq
    0\right\}\,.
\end{equation}
Here, $K_{T,\mu}$ is the multiplication operator in momentum space
$$
K_{T,\mu} = \left( p^2 -\mu\right) \frac { e^{(p^2-\mu)/T}+1
}{e^{(p^2-\mu)/T}-1}\, .
$$
Note that $K_{T,\mu}\geq 2T \geq 0$, and $\lim_{T\to 0} K_{T,\mu} = |p^2-\mu|$. Observe that Eq.~(\ref{deftc}) characterizes the critical temperature for the {\it nonlinear} BCS equation in terms of the spectrum of {\it linear} operators.

We assume that $\mu>0$ henceforth. For small coupling $\lambda$, the
critical temperature is determined by the behavior of the potential $V(x)$ on
the Fermi sphere $\Omega_\mu$, the sphere in momentum space with
radius $\sqrt{\mu}$.  We denote the uniform (Lebesgue) measure on $\Omega_\mu$
by $d\omega$.

Let $\V_\mu: \, L^2(\Omega_\mu) \to L^2(\Omega_\mu)$ be the
self-adjoint operator
\begin{equation}\label{defvm}
  \big(\V_\mu u\big)(p) =
  \frac 1{(2\pi)^{3/2}} \frac 1{\sqrt{\mu}}\int_{\Omega_\mu}\hat V(p-q) u(q) \,d\omega(q)\,.
\end{equation}
Since $V\in L^1(\R^3)$ by assumption, $\hat V(p)$ is a bounded
continuous function, and hence $\V_\mu$ is a Hilbert-Schmidt
operator. It is, in fact, trace class, and its trace equals
$\frac{\sqrt{\mu}}{2\pi^2} \int_{\R^3} V(x)dx$.  Let
$$
e_{\mu}= \infspec \V_\mu
$$
denote the infimum of the spectrum of $\V_\mu$.  Since $\V_\mu$ is
compact, we have $e_\mu\leq 0$. We note that operators of the form
(\ref{defvm}) have appeared in related works on bound states and
scattering properties of pseudo-differential operators, see
\cite{BY,LSW}.

In \cite{FHNS}, it was shown that when $e_\mu<0$ the asymptotic
behavior of $T_c$ as $\lambda$ tends to zero is, to leading order, given by
\begin{equation}\label{tc1}
  \lim_{\lambda\to 0} \lambda \ln \frac\mu T_c = -\frac{1}{e_{\mu}}\,.
\end{equation}
That is, $T_c \sim \exp(-1/\lambda|e_\mu|)$ for small $\lambda$.
In the following, we shall derive the second order correction,
i.e., we will compute the constant in front of the exponentially small
term in $T_c$. For this purpose, we define an operator $\W_\mu$ on
$L^2(\Omega_\mu)$ via its quadratic form
\begin{align}\nonumber
  \langle u | \W_\mu |u\rangle = \int_{0}^\infty d|p| & \left( \frac
    {|p|^2}{\big||p|^2-\mu\big|} \left[ \int_{\S^2} d\Omega \left(
        |\hat\varphi(p)|^2 -
        |\hat\varphi(\sqrt\mu p/|p|)|^2 \right)\right] \right. \\
  \label{defW} & \quad \left. + \int_{\S^2} d\Omega\,
    |\hat\varphi(\sqrt\mu p/|p|)|^2\right) \,.
\end{align}
Here, $\hat\varphi(p) = (2\pi)^{-3/2} \int_{\Omega_\mu} \hat V(p-q)
u(q) d\omega(q)$, and $(|p|,\Omega)\in \R_+\times \S^2$ denote
spherical coordinates for $p\in\R^3$. We note that since $V\in
L^1(\R^3)$, $\int_{\S^2} d\Omega\,|\hat\varphi(p)|^2$ is Lipschitz
continuous in $|p|$ for any $u\in L^2(\R^3)$. (See Eq.~(\ref{lips})
below.) Hence the radial integral in (\ref{defW}) is well-defined,
even in the vicinity of $|p|\sim \sqrt\mu$. For large $|p|$, the
integral converges because $V\in L^{3/2}(\R^3)$. We shall, in fact, see that
the operator $\W_\mu$ is of Hilbert-Schmidt class (see the proof of
Theorem~\ref{constant} in Section~\ref{proof:thm1}).

For $\lambda >0$, let
\begin{equation}\label{defB}
  \B_\mu = \lambda \frac \pi {2\sqrt \mu} \V_\mu 
- \lambda^2 \frac{\pi}{2\mu} \W_\mu\,,
\end{equation}
and let $b_\mu(\lambda)$ denote its ground state energy,
\begin{equation}\label{defbm}
  b_\mu(\lambda) = \infspec \B_\mu \,.
\end{equation}
We note that if $e_\mu<0$, then also $b_\mu(\lambda)< 0$ for small
$\lambda$. In fact, if the eigenfunction corresponding to the lowest
eigenvalue $e_\mu$ of $\V_\mu$ is unique and given by $u\in
L^2(\Omega_\mu)$, then
\begin{equation}\label{bmexpl}
b_\mu(\lambda) = \langle u|\B_\mu|u\rangle + O(\lambda^3) = \lambda
\frac{ \pi e_\mu}{2\sqrt\mu} - \lambda^2 \frac{\pi \langle u|
  \W_\mu|u\rangle}{2\mu} + O(\lambda^3)\,.
\end{equation}
In the degenerate case, this formula holds if one chooses $u$ to be
the eigenfunction of $\V_\mu$ that yields the largest value $\langle
u|\W_\mu|u\rangle$ among all such (normalized) eigenfunctions.

With the aid of $b_\mu(\lambda)$, we can now state our first main
result concerning the asymptotic behavior of the critical temperature
$T_c$ for small $\lambda$.

\begin{Theorem}[{\bf Critical Temperature}]\label{constant}
  Let $V\in L^1(\R^3)\cap L^{3/2}(\R^3)$ and let $\mu>0$.  Assume that
  $e_\mu = \infspec \V_\mu <0$, and let $b_\mu(\lambda)$ be defined in
  (\ref{defbm}).  Then the critical temperature $T_c$ for the BCS
  equation, given in Eq.~(\ref{deftc}), is strictly positive and satisfies
  \begin{equation}\label{themeq}
    \lim_{\lambda \to 0} \left(\ln\left(\frac\mu {T_c}\right) + 
      \frac {\pi}{2 \sqrt{\mu}\, b_\mu(\lambda)}\right) = 2 - \gamma - \ln(8/\pi)\,.
  \end{equation}
  Here, $\gamma\approx 0.577$ denotes Euler's constant.
\end{Theorem}

The Theorem says that, for small $\lambda$,
\begin{equation}\label{formula}
  T_c \sim \mu \frac{8 e^{\gamma-2}}{\pi} e^{\pi/(2 \sqrt{\mu} b_\mu(\lambda))} \,.
\end{equation}
Eq.~(\ref{bmexpl}) can be reformulated as 
\begin{equation}\label{aab}
\frac \pi{2\sqrt\mu\, b_\mu(\lambda)} = \frac {1}{\lambda e_\mu} +
\frac {\langle u|\W_\mu|u\rangle}{ \sqrt\mu\, e_\mu^2} +O(\lambda)
\end{equation}
as $\lambda \to 0$, where $u$ is an eigenfunction of $\V_\mu$ in (\ref{defvm}) with
eigenvalue $e_\mu<0$. We note that for radial potentials $V(x)$, the eigenfunction $u$ corresponding to the
lowest eigenvalue $e_\mu$ of $\V_\mu$  will be an
eigenfunction of the angular momentum. It need not have zero angular
momentum, however, but can, in principle, have arbitrarily high
angular momentum depending on the details of $V(x)$.

Because of (\ref{aab}),  Theorem~\ref{constant} could alternatively
be formulated as
$$
\lim_{\lambda \to 0} \left(\ln\left(\frac\mu {T_c}\right) + \frac
  {1}{\lambda e_\mu}\right) = 2 - \gamma - \ln(8/\pi)- \frac {\langle
    u|\W_\mu|u\rangle}{ \sqrt\mu\, e_\mu^2}\,.
$$
It is natural, however, to absorb the last term into the definition of
$b_\mu(\lambda)$, as we do here, since $b_\mu(\lambda)$ can be
interpreted as a (renormalized) effective scattering length of
$2\lambda V(x)$ (in second order Born approximation) for particles
with momenta on the Fermi sphere. In fact, if $V$ is {\it radial} and
$\int_{\R^3} V(x)dx < 0$, it is not difficult to see that for small enough
$\mu$ the (unique) eigenfunction corresponding to the lowest
eigenvalue $e_\mu$ of $\V_\mu$ is the constant function
$u(p)=(4\pi\mu)^{-1/2}$. (See Section~2.1 in \cite{FHNS}.) For this
$u$, we have
$$
\lim_{\mu\to 0} \langle u|\B_\mu|u\rangle = (\lambda/4\pi)\int_{\R^3} V(x) dx
- (\lambda/4\pi)^2 \int_{\R^6} \frac{V(x) V(y)}{|x-y|}dxdy \equiv
a_0(\lambda)\,.
$$ 
Here, $a_0(\lambda)$ equals the {\it scattering length} of $2\lambda V$ in
second order Born approximation. Assuming a certain decay rate of $V$
at infinity we can, in fact, estimate the difference between $\langle
u|\B_\mu|u\rangle$ and $a_0(\lambda)$.

\begin{Proposition}[{\bf Relation to Scattering Length}]\label{balem}
  Let $V \in L^1(\R^3)\cap L^{3/2}(\R^3)$, $\mu>0$, and let
  $u(p)=(4\pi\mu)^{-1/2}$ be the constant function on the sphere
  $\Omega_\mu$.  Let $0\leq \beta < 2$, and assume that $\int_{\R^3} |V(x)|
  |x|^{\beta} dx < \infty$. Then
$$
\lim_{\mu\to 0} \frac 1{\sqrt\mu^\beta}\big( \langle
  u|\B_\mu|u\rangle - a_0(\lambda)\big) =0\,.
$$
\end{Proposition}

In particular, if $\beta \geq 1$, this implies that
\begin{equation}\label{ba}
  \lim_{\mu\to 0} \frac 1{\sqrt\mu}\left(\frac 1{\langle u|\B_\mu|u\rangle} - \frac 1
    {a_0(\lambda)}\right)=0\,.
\end{equation}
As explained above, for radial potentials $V$ with $\int_{\R^3} V(x)dx
<0$ the eigenfunction corresponding to the lowest eigenvalue of
$\V_\mu$ in (\ref{defvm}) is the constant function $u(p)=(4\pi
\mu)^{-1/2}$ for small $\mu$, and thus $b_\mu(\lambda) = \langle
u|\B_\mu|u\rangle + O(\lambda^3)$ for small enough $\mu$ in this
case. Hence (\ref{themeq}) holds, with $b_\mu(\lambda)$ replaced by
$\langle u|\B_\mu|u\rangle$.  If, in addition, $\int_{\R^3} |V(x)||x|dx
<\infty$, one can use (\ref{ba}) to replace $\langle
u|\B_\mu|u\rangle$ by $a_0(\lambda)$, and arrive at the approximation
$$T_c \approx \mu \frac{8 e^{\gamma-2}}{\pi} e^{\pi/(2 \sqrt{\mu} a_0(\lambda))}
$$
for the critical temperature for small $\lambda$ {\it and} small
$\mu$.  This expression is well-known in the physics literature, see,
e.g., \cite{gorkov}. It is valid beyond the small coupling
approximation \cite{HS08}.  We point out, however, that our formula
(\ref{formula}) is much more general since it holds for any value of
$\mu>0$.

\subsection{Energy Gap at Zero Temperature}

Consider now the zero temperature case $T=0$. In this case, it is
natural to formulate a functional depending only on $\alpha$ instead
of $\gamma$ and $\alpha$. In fact, for $T=0$  the
optimal choice of $\gamma(p)$ in $\F$ for given $\hat\alpha(p)$ is clearly
\begin{equation}\label{gal}
  \gamma(p) = \left\{ \begin{array}{ll}
      \half (1+\sqrt{1-4|\hat\alpha(p)|^2}) & {\rm for\ } p^2< \mu \\
      \half (1-\sqrt{1-4|\hat\alpha(p)|^2}) & {\rm for\ } p^2>\mu
    \end{array}\right.\,.
\end{equation}
Subtracting an unimportant constant, this leads to the {\it zero temperature
BCS functional}
\begin{equation}\label{deffa}
  \F_0(\alpha)
  =\frac 12 \int_{\R^3} |p^2-\mu|\left(1-\sqrt{1-4|\hat\al(p)|^2}\right)dp+ \lambda \int_{\R^3}
  V(x)|\alpha(x)|^2\,dx\,.
\end{equation}

Defining $\Delta$ as in (\ref{defdelta}) and inserting (\ref{gal}),
the relation between $\Delta$ and $\alpha$ at $T=0$ is
\begin{equation}\label{dat0}
\Delta(p) = 2 \frac{|p^2-\mu| \hat \alpha(p)}{\sqrt{1-4|\hat\alpha(p)|^2}}\,.
\end{equation}
The variational equation satisfied by a minimizer of (\ref{deffa}) is then
\begin{equation}\label{bcset}
  \Delta(p) = -\frac \lambda{(2\pi)^{3/2}} \int_{\R^3} \hat V(p-q)
  \frac{\Delta(q)}{E(q)} \, dq\,.
\end{equation}
This is simply the BCS equation (\ref{bcse}) at $T=0$.  For a solution
$\Delta$, the {\it energy gap} $\Xi$ is defined as
\begin{equation}\label{defxi}
\Xi = \inf_p E(p) = \inf_p \sqrt{(p^2-\mu)^2 + |\Delta(p)|^2}\,.
\end{equation}
It has the interpretation of an energy gap in the corresponding 
second-quantized BCS Hamiltonian (see, e.g., \cite{MR} or the appendix in
\cite{HHSS}.)

One of the difficulties involved in evaluating $\Xi$ is the potential
non-uniqueness of minimizers of (\ref{deffa}), and hence
non-uniqueness of solutions of the BCS gap equation (\ref{bcset}). The
gap $\Xi$ may depend on the choice of $\Delta$ in this case. For
potentials $V$ with non-positive Fourier transform, however, we can
prove the uniqueness of $\Delta$ and, in addition, we are able to
derive its precise asymptotics as $\lambda\to 0$. This is the content
of Lemma~\ref{Deltach} below. In particular, this allows us the derive
an expression for $\Xi$ in the limit $\lambda\to 0$, which is stated
in Theorem~\ref{gap}.

We will restrict our attention to radial potentials $V$ with
non-positive Fourier transform in the following.  We also assume that
$\hat V(0) = (2\pi)^{-3/2} \int V(x) dx < 0$.  It is easy to see that
$e_\mu = \infspec \V_\mu <0$ in this case, and that the (unique)
eigenfunction corresponding to this lowest eigenvalue of $\V_\mu$ is
the constant function.
Under these assumptions on $V$, we have the following asymptotic
behavior of the energy gap $\Xi$ as $\lambda \to 0$.

\begin{Theorem}[{\bf Energy gap}]\label{gap}
  Assume that $V\in L^1(\R^3)\cap L^{3/2}(\R^3)$ is radial, with $\hat
  V(p)\leq 0$ and $\hat V(0)<0$. Then there is a unique minimizer (up
  to a constant phase) of the BCS functional (\ref{deffa}) at
  $T=0$. The corresponding energy gap,
$$
\Xi = \inf_p \sqrt{ (p^2-\mu)^2 + |\Delta(p)|^2}\,,
$$
is strictly positive, and satisfies
\begin{equation}\label{themeq2}
  \lim_{\lambda \to 0} \left(\ln\left(\frac\mu \Xi \right) + 
    \frac {\pi}{2 \sqrt{\mu}\, b_\mu(\lambda)}\right) = 2 - \ln(8)\,.
\end{equation}
Here, $b_\mu(\lambda)$ is defined in (\ref{defbm}).
\end{Theorem}

The Theorem says that, for small $\lambda$,
$$
\Xi \sim \mu \frac{8}{e^2} e^{\pi/(2 \sqrt{\mu} b_\mu(\lambda))} \,.
$$
In particular, in combination with Theorem~\ref{constant}, we obtain the following corollary of Theorem~\ref{gap}.

\begin{corollary}[{\bf Universal Ratio}]
  Under the same assumptions as in Theorem~\ref{gap}, the ratio of the
  energy gap $\Xi$ and the critical temperature $T_c$ satisfies
$$
\lim_{\lambda\to 0} \frac{ \Xi}{T_c} = \frac \pi{e^\gamma} \approx
1.7639\,.
$$
\end{corollary}

That is, the ratio of the energy gap $\Xi$ and the critical
temperature $T_c$ tends to a universal constant as $\lambda\to 0$,
independently of $V$ and $\mu$. This property has been observed before
for the original BCS model with rank one interaction \cite{BCS,MR}, and
in the low density limit for more general interactions \cite{gorkov}
under additional assumptions. Our analysis shows that it is valid in
full generality at small coupling $\lambda \ll 1$.

We remark that although Theorem~\ref{gap} can be expected to hold under
weaker assumptions on the potential $V$ than the ones considered here,
stronger assumptions than merely $e_\mu<0$ (as in Theorem~\ref{constant})
are needed for positivity of $\Xi$. In particular, if $\Delta$ has
non-zero angular momentum, $\Xi$ will, in general, vanish.

\section{Proof of Theorem~\ref{constant}}\label{proof:thm1}

For a (not necessarily sign-definite) potential $V(x)$ let us use the
notation
\begin{equation*}
  V(x)^{1/2} = ({\rm sgn\,} V(x)) |V(x)|^{1/2} \,.
\end{equation*}
The Birman-Schwinger principle (see Lemma~1 in \cite{FHNS}) implies
that the critical temperature $T_c$ in (\ref{deftc}) is determined by
the fact that for this value of $T$ the smallest eigenvalue of
\begin{equation}\label{defofbt}
  B_T = \lambda V^{1/2}K_{T,\mu}^{-1}|V|^{1/2}
\end{equation}
equals $-1$. Note that although $B_T$ is not self-adjoint, it has a real
spectrum.

Let $\dig: L^1(\R^3) \to L^2(\Omega_\mu)$ denote the (bounded)
operator that maps $\psi\in L^1(\R^3)$ to the Fourier transform of
$\psi$, restricted to the sphere $\Omega_\mu$. Since $V\in L^1(\R^3)$,
multiplication by $|V|^{1/2}$ is a bounded operator from $L^2(\R^3)$ to
$L^1(\R^3)$, and hence $\dig |V|^{1/2}$ is a bounded operator from
$L^2(\R^3)$ to $L^2(\Omega_\mu)$. Let
$$
m_\mu(T) = \max\left\{ \frac 1{4\pi \mu} \int_{\R^3} \left( \frac 1{K_{T,\mu}(p)}
  - \frac 1{p^2}\right) dp\, , 0\right\} \,,
$$
and let
\begin{equation}\label{defmt}
  M_T = K_{T,\mu}^{-1} - m_\mu(T) \dig^* \dig\,.
\end{equation}
It was shown in \cite[Lemma~2]{FHNS} that $ V^{1/2} M_T |V|^{1/2}$
is a Hilbert-Schmidt operator on $L^2(\R^3)$, and its Hilbert Schmidt
norm is bounded uniformly in $T$. (In fact, in \cite{FHNS} a slightly
different definition of $m_\mu(T)$ was used, but it differs from ours
only by a term that is uniformly bounded in $T$.) In particular, the
singular part of $B_T$ as $T\to 0$ is determined entirely by
$V^{1/2}\dig^* \dig |V|^{1/2}$.

Since $V^{1/2} M_T |V|^{1/2}$ is uniformly bounded, we can choose
$\lambda$ small enough such that $\sup_{T>0} \| V^{1/2} M_T|V|^{1/2}
\| < 1/\lambda$. Then $1+\lambda V^{1/2} M_T |V|^{1/2}$ is invertible for any $T>0$,
and we can thus write $1+ B_T$ as
\begin{align}\label{1ba}
  1+ B_T &= 1+ \lambda V^{1/2} \left( m_\mu(T) \dig^* \dig +
    M_T\right) |V|^{1/2} \\ \nonumber &= \left(1+ \lambda V^{1/2} M_T
    |V|^{1/2} \right) \left( 1 + \frac{\lambda m_\mu(T)}{1+ \lambda
      V^{1/2} M_T |V|^{1/2}} V^{1/2} \dig^* \dig |V|^{1/2}\right)\,.
\end{align}
Then $B_T$ having an eigenvalue $-1$  is equivalent to
\begin{equation}\label{a1}
  \frac{\lambda m_\mu(T)}{1+ \lambda V^{1/2} M_T |V|^{1/2}} V^{1/2}
  \dig^* \dig |V|^{1/2}
\end{equation}
having an eigenvalue $-1$. The operator in (\ref{a1}) is
isospectral to the selfadjoint operator
\begin{equation}\label{b1}
\dig |V|^{1/2} \frac {  \lambda m_\mu(T) }{1+ \lambda V^{1/2} M_T
    |V|^{1/2}} V^{1/2} \dig^*\,,
\end{equation}
acting on $L^2(\Omega_\mu)$. 

At $T=T_c$, $-1$ is the smallest eigenvalue of $B_T$, hence (\ref{a1})
and (\ref{b1}) have an eigenvalue $-1$ for this value of
$T$. Moreover, we can conclude that $-1$ is actually the {\it smallest}
eigenvalue of (\ref{a1}) and (\ref{b1}) in this case. For, if there
were an eigenvalue less than $-1$, we could increase $T$ and, by
continuity (and the fact that $m_\mu(T)$ is monotone decreasing and
goes to zero as $T\to \infty$), find some $T>T_c$ for which there is
an eigenvalue $-1$. Using (\ref{1ba}), this would contradict the fact
that $B_T$ has no eigenvalue $-1$ for $T>T_c$.

Note that $\dig V \dig^* = \sqrt\mu\, \V_\mu$ defined in
(\ref{defvm}). By assumption, $e_\mu = \infspec \V_\mu$ is strictly
negative.  Since for $T=T_c$ the smallest eigenvalue of (\ref{b1})
equals $-1$, it follows immediately that
$$
\lim_{\lambda \to 0} \lambda m_\mu(T_c) =- \frac 1{\infspec \dig V
  \dig^*} = - \frac 1 {\sqrt\mu \, e_\mu}\,.
$$
Together with the asymptotic behavior $m_\mu(T) \sim \mu^{-1/2}
\ln(\mu/T)$ as $T\to 0$, this implies that the leading order behavior
of $\ln (\mu/T_c)$ as $\lambda\to 0$ is given by (\ref{tc1}), as was
proved in \cite{FHNS}.

To obtain the next order, we employ first order perturbation theory.
Since $\dig V \dig^*$ is compact and $\infspec \dig V \dig^*<0$ by
assumption, first order perturbation theory implies that
\begin{equation}\label{deno}
  m_\mu(T_c) = \frac {-1}{ \lambda \langle u| \dig V \dig^*| u\rangle -
    \lambda ^2 \langle u| \dig V M_{T_c} V \dig^*| u\rangle + O(\lambda^3)}\,,
\end{equation}
where $u$ is the (normalized) eigenfunction corresponding to the lowest eigenvalue
of $\dig V \dig^*$. (In case of degeneracy, one has to the choose the
$u$ that minimizes the $\lambda^2$ term in the denominator of
(\ref{deno}) among all such eigenfunctions.)

Eq.~(\ref{deno}) is an implicit equation for $T_c$. Since $\dig V M_T
V\dig^*$ is uniformly bounded and $T_c\to 0$ as $\lambda \to 0$, we
have to evaluate the limit of $\langle u| \dig V M_T V \dig^*| u\rangle$ as $T\to
0$. For this purpose, let $\varphi = V \dig^* u$. Then
\begin{align}\nonumber 
  &\langle u| \dig V M_T V \dig^* |u\rangle \\ \label{comeq}
& = \int_{\R^3} \frac 1{K_{T,\mu}(p)}
  |\hat\varphi(p)|^2 \, dp - m_\mu(T) \int_{\Omega_\mu}
  |\hat\varphi(p)|^2 \, d\omega(p) \\ \nonumber & = \int_{\R^3}
  \left(\frac 1{K_{T,\mu}(p)} \left[ |\hat\varphi(p)|^2
      -|\hat\varphi(\sqrt\mu p/|p|)|^2 \right] + \frac 1{p^2}
    |\hat\varphi(\sqrt\mu p/|p|)|^2 \right) dp\,.
\end{align}
Recall that $K_{T,\mu}(p)=|p^2-\mu|/\tanh(|p^2-\mu|/2T)$, which
converges to $|p^2-\mu|$ as $T\to 0$. We claim that the spherical
average of $|\hat\varphi(p)|^2$ is Lipschitz continuous. In fact,
\begin{align}\label{lips}
  &(2\pi)^3 \int_{\S^2} d\Omega\, |\hat\varphi(p)|^2 \\ \nonumber & =
  \sqrt \frac 2\pi \int_{\R^6} dxdy\, \frac{\sin( |p||x-y|)}{|p||x-y|}
  V(x) V(y) \int_{\Omega_\mu} e^{iqx} u(q) d\omega(q)
  \int_{\Omega_\mu} e^{-iry} \overline{u(r)} d\omega(r)\,,
\end{align}
and hence Lipschitz continuity in $|p|$ follows from the fact that
$V\in L^1(\R^3)$, $u\in L^2(\Omega_\mu)$ and $|\sin(a)/a-\sin(b)/b|
\leq C |a-b|/(a+b)$ for $a>0$, $b>0$ and some constant $C>0$.

Using Lipschitz continuity, it is then easy to see that one can
interchange the limit and the radial integral over $|p|$, and hence
obtain
\begin{equation}\label{deno2}
  \lim_{T\to 0} \langle u| \dig V M_T V \dig^*| u\rangle  = \langle u| \W_\mu |u \rangle\,,
\end{equation}
with $\W_\mu$ defined in (\ref{defW}). Moreover, this convergence is
uniform in $u\in L^2(\Omega_\mu)$. Since $\dig V M_T V \dig^*$ is uniformly bounded in the 
Hilbert-Schmidt norm, this also shows that $\W_\mu$ is a
Hilbert-Schmidt operator, as claimed in Section~\ref{subsec:ct}. In
particular, combining (\ref{deno}) and (\ref{deno2}), we have thus
shown that
\begin{equation}\label{denof}
  \lim_{\lambda\to 0} \left( m_\mu(T_c) + \frac 1{\infspec \left
        (\lambda \sqrt \mu\, \V_\mu - \lambda^2 \W_\mu\right)} \right) =
  0\,.
\end{equation}

It remains to calculate $m_\mu(T)$.  Recall that
$K_{T,\mu}(p)=|p^2-\mu|/\tanh(|p^2-\mu|/2T)$. We claim the following. 

\begin{Lemma}\label{calcm}
  As $T\to 0$,
  \begin{equation}\label{lemresult}
    m_\mu(T) = \frac 1{4\pi \mu} \int_{\R^3}\left( \frac 1
      {K_{T,\mu}(p) } -\frac 1{p^2}\right) dp  = \frac 1{\sqrt\mu}
\left( \ln \frac \mu T + \gamma-2 + \ln\frac 8\pi  + o(1)\right) \,.
  \end{equation}
  Here, $\gamma\approx 0.5772$ is Euler's constant.
\end{Lemma}

\begin{proof}
  By splitting the integral into two parts according to $p^2\leq \mu$
  and $p^2\geq \mu$, and changing variables from $p^2-\mu$ to $-t$ and
  $t$, respectively, we see that
  \begin{align}\label{split1}
    m_\mu(T) & = \frac 1{2\mu} \int_0^\mu\left( \frac{\sqrt{\mu-t}}{t}
      \tanh(t/2T) - \frac{1}{\sqrt{\mu-t}}\right) dt \\ & \nonumber
    \quad + \frac 1{2\mu} \int_0^{\infty} \left(\frac{\sqrt{\mu+t}}{t}
      \tanh(t/2T) - \frac 1{\sqrt{\mu+t}}\right) dt \,.
  \end{align}
  Using dominated convergence,
  \begin{align}\nonumber
    &\lim_{T\to 0} \int_\mu^{\infty} \left(\frac{\sqrt{\mu+t}}{t}
      \tanh(t/2T) - \frac 1{\sqrt{\mu+t}}\right) dt \\ \nonumber &=
    \int_\mu^{\infty} \left(\frac{\sqrt{\mu+t}}{t} - \frac
      1{\sqrt{\mu+t}}\right) dt = 2 \sqrt\mu \ln\left(1+\sqrt 2\right)\,.
  \end{align}

For the integrals for $0\leq t\leq \mu$, we use
$$
\int_0^\mu \left( \frac 1{\sqrt{\mu-t}} + \frac 1{\sqrt{\mu+t}}\right)
dt = 2 \sqrt{2\mu}\,.
$$
Moreover, using again dominated convergence, one sees that
$$
\lim_{T\to 0} \int_0^\mu \frac {\sqrt{\mu-t}-\sqrt{\mu}}t
\tanh(t/2T)\, dt = \int_0^\mu \frac {\sqrt{\mu-t}-\sqrt{\mu}}t dt = 2
\sqrt{\mu} \left( \ln 2 -1\right)\,.
$$
Similarly,
$$
\lim_{T\to 0} \int_0^{\mu} \frac{\sqrt{\mu+t}-\sqrt{\mu}}t
\tanh(t/2T)\, dt = 2 \sqrt\mu\left( \ln 2 -1 + \sqrt 2 - \ln
  \left(1+\sqrt 2\right)\right)\,.
$$

In order to calculate the remaining integral $\int_0^\mu
t^{-1} \tanh(t/2T) dt$, we split the hyperbolic tangent into two parts,
$$
\tanh(t/2T) = \left(1-\exp(-t/T)\right) - \exp(-t/T) \tanh(t/2T)\,.
$$
Using partial integration,
\begin{align}\nonumber
  \int_0^\mu \frac{1-\exp(-t/T)}{t} dt & \nonumber = \int_0^{\mu/T}
  \frac {1-\exp(-t)}{t} dt \\ &= \ln \frac\mu T \left(
    1-\exp(-\mu/T)\right) - \int_0^{\mu/T} \ln (t)\exp(-t) \, dt\,.
\end{align}
Since $-\int_0^\infty \ln(t)\exp(-t)dt$ equals Euler's constant
$\gamma\approx 0.5772$ \cite[4.331.1]{gr}, we get
$$
\lim_{T\to 0} \left( \int_0^\mu \frac{1-\exp(-t/T)}{t} dt -
  \ln\frac\mu T\right) = \gamma\,.
$$
Finally, we have to evaluate
$$
\int_0^\mu \frac{\tanh(t/2T)}{t} \exp(-t/T) \, dt = \int_0^{\mu/T}
\frac{\tanh(t/2)}t \exp(-t) \, dt\,.
$$
In the limit $T\to 0$, this becomes \cite[3.411.28]{gr}
$$
\int_0^{\infty} \frac{\tanh(t/2)}t \exp(-t) \, dt = \ln \frac \pi 2\,.
$$
Collecting all the terms, we arrive at (\ref{lemresult}).
\end{proof}

Theorem~\ref{constant} follows immediately from (\ref{denof}) and
(\ref{lemresult}), recalling the definition of $b_\mu(\lambda)$ in
(\ref{defB}) and (\ref{defbm}).

\section{Proof of Proposition~\ref{balem}}

For $u(p)=(4\pi\mu)^{-1/2}$ the (normalized) constant function in $L^2(\Omega_\mu)$, 
we have
\begin{align}\nonumber
\langle u|\B_\mu|u\rangle -a_0(\lambda) &= \lambda \sqrt\frac\pi 2
  \int_{\Omega_\mu\times \Omega_\mu} \left(\hat V(p-q) - \hat
    V(0)\right) \, \frac{d\omega(p)}{4\pi \mu} \,
  \frac{d\omega(q)}{4\pi\mu} \\ \nonumber & \quad - \frac{
    \lambda^2}{4\pi} \int_{\R^3} \left( \frac 1{|p^2- \mu|}- \frac 1
    {p^2}\right)\left( |\psi_\mu(p)|^2 - |\psi_\mu(\sqrt\mu
    p/|p|)|^2\right) dp \\ \label{ppo} & \quad +
  \frac{\lambda^2}{4\pi} \int_{\R^3} \frac 1 {p^2} \left(
    |\psi_0(p)|^2 - |\psi_\mu(p)|^2\right) dp\,,
\end{align}
where $\psi_\mu(p) = \int_{\Omega_\mu} \hat V (p -q ) d\omega_\mu(q) /(4\pi\mu)$
and $\psi_0(p) = \hat V(p)$.

Consider first the term linear in $\lambda$. It is given by
$$
\frac \lambda{4\pi} \int_{\R^3} V(x) \left( \frac{\sin^2(\sqrt\mu
    |x|)}{\mu |x|^2} - 1\right)dx \,.
$$
For $0\leq \beta< 2$, $(\sqrt\mu |x|)^{-\beta}(\sin^2(\sqrt\mu
|x|)/(\mu |x|^2) - 1)$ is a function that is bounded independently of
$\mu$, and goes to zero as $\mu\to 0$ for every $x$. Hence, if $\int
|V(x)| |x|^\beta dx < \infty$, it follows from dominated convergence
that
$$
\lim_{\mu \to 0} \frac 1{\sqrt{\mu}^\beta} \int_{\R^3} V(x) \left(
  \frac{\sin^2(\sqrt\mu |x|)}{\mu |x|^2} - 1\right)dx =0 \,.
$$

The second term in (\ref{ppo}) can be rewritten as
\begin{align}\nonumber
  &\lambda^2 \int_{\R^6} V(x) \frac{\sin(\sqrt\mu|x|)}{\sqrt{\mu}|x|}
  V(y) \frac{\sin(\sqrt\mu|y|)}{\sqrt{\mu}|y|} \int_0^\infty \left(
    \frac{p^2}{|p^2-\mu|} - 1\right) \\ \label{set} &\qquad \qquad
  \times\left(
    \frac{\sin(p|x|)}{p|x|}-\frac{\sin(\sqrt\mu|x|)}{\sqrt{\mu}|x|}\right)
  \left(
    \frac{\sin(p|y|)}{p|y|}+\frac{\sin(\sqrt\mu|y|)}{\sqrt{\mu}|y|}\right)
  dp\, dx\, dy \,.
\end{align}
The last term in the integrand is bounded by $2$, while the second
term is bounded by $C|p-\sqrt\mu| \min\{|x|,1/(p+\sqrt\mu)\}$, since
$|\sin(a)/a-\sin(b)/b| \leq C |a-b| \min\{1,1/(b+a)\}$. Since $0\leq
\beta<2$, we can estimate $ \min\{|x|,1/(p+\sqrt\mu)\} \leq
|x|^{\beta/2} (p+\sqrt\mu)^{\beta/2-1}$. Hence we conclude that
(\ref{set}) is bounded by
\begin{align}\nonumber
  & 2 \lambda^2 C \|V\|_1 \int_{\R^3} |V(x)| |x|^{\beta/2} dx\,
  \int_0^\infty \left( \frac{p^2}{|p^2-\mu|} - 1\right) \frac
  {|p-\sqrt\mu|}{(p+\sqrt{\mu})^{1-\beta/2}}dp \\ \nonumber & = 2
  \lambda^2 C \sqrt{\mu}^{1 + \beta/2} \|V\|_1 \int_{\R^3} |V(x)|
  |x|^{\beta/2} dx \, \int_0^\infty \left( \frac{p^2}{|p^2-1|} -
    1\right) \frac {|p-1|}{(p+1)^{1-\beta/2}}dp \,.
\end{align}
Since $1+\beta/2 > \beta$ this yields a bound of the desired form.

Finally, the last term in (\ref{ppo}) is given by
$$
\left(\frac \lambda {4\pi}\right)^2 \int_{\R^6} \frac {V(x)
  V(y)}{|x-y|} \left( 1 - \frac{\sin(\sqrt\mu |x|)}{\sqrt\mu |x|}
  \frac{\sin(\sqrt\mu |y|)}{\sqrt\mu |y|}\right) dx dy\,.
$$ 
The Hardy-Littlewood-Sobolev inequality \cite{LL} and H\"older's
inequality imply that 
$$
\int \frac{V(x)V(y)}{|x-y|} |x|^{\beta/2}
|y|^{\beta/2} dx dy \leq C \big\| V|\,\cdot\,|^{\beta/2}\big\|_{6/5}^2 \leq C
\|V\|_{3/2} \int |V(x)||x|^{\beta} dx
$$
for some positive constant $C$. Moreover, since for $0\leq \beta<2$ the
expression $(\mu |x| |y|)^{-\beta/2} \left( 1 - \sin(\sqrt\mu
  |x|)\sin(\sqrt\mu |y|)/(\mu |x| |y|)\right)$ is uniformly bounded
and goes to zero pointwise as $\mu\to 0$, it follows again from
dominated convergence that
$$
\lim_{\mu\to 0} \frac 1{\sqrt\mu^\beta} \int_{\R^6} \frac {V(x)
  V(y)}{|x-y|} \left( 1 - \frac{\sin(\sqrt\mu |x|)}{\sqrt\mu |x|}
  \frac{\sin(\sqrt\mu |y|)}{\sqrt\mu |y|}\right) dx dy = 0\,.
$$
This proves Proposition~\ref{balem}.

\section{Proof of Theorem~\ref{gap}}\label{sec:gap}

We start by showing that the minimizer of the BCS functional
(\ref{deffa}) at $T=0$ is unique under the assumption that $\hat V$ is
non-positive and $\hat V(0) < 0$.  Note that if, in addition, $V$ is
radial, this necessarily implies that also the minimizer has to be
radial.

\begin{Lemma}\label{uniqueness}
  Let $V \in L^1 \cap L^{3/2}$ and assume that $\hat V \leq 0$ and
  $\hat V(0)<0$. Then the BCS functional $\F_0$ at $T=0$, defined in
  (\ref{deffa}), has a unique minimizer $\alpha$ (modulo a constant
  phase), whose Fourier transform $\hat \alpha$ is strictly positive.
\end{Lemma}

In particular, this implies that $\Delta$ is strictly positive. From
the gap equation (\ref{bcset}) one easily concludes that $\Delta$ is
also continuous, and hence $\Xi>0$.

\begin{proof}
  The existence of a minimizer was shown in \cite{HHSS}. Moreover,
  since $e_\mu = \infspec \V_\mu<0$ under our assumptions on $V$, the
  critical temperature $T_c$ is strictly positive, and hence a
  minimizer of $\F_0$ is necessarily not identically zero.

  Note that since $\hat V \leq 0$,
  \begin{equation}\label{sq}
    \int_{\R^6} \overline{\hat\alpha(p)} \hat V(p-q) \hat\alpha(q) \, dpdq \geq 
\int_{\R^6} |\hat\alpha(p)| \hat V(p-q)| \hat\alpha(q)|\, dpdq \,.
  \end{equation}
  Hence, if $\hat\alpha(p)$ is a minimizer of $\F_0$, so is $|\hat \alpha(p)|$.

  Assume now that there are two different minimizers $f\neq
  g$, both with nonnegative Fourier transform.  Since $t \mapsto 1 -
  \sqrt{1 - 4 t}$ is strictly convex for $0\leq t \leq 1/2$ we 
  see that $\psi = \frac 1{\sqrt{2}} f + i \frac 1{\sqrt{2}} g$,
  satisfies
$$ \F_0 (\psi)  < \half \F_0(f) + \half \F_0(g)\,.$$
This is a contradiction to $f,g$ being distinct minimizers, and hence
$f=g$. In particular, the absolute value of a minimizer is unique.

Let $\hat\alpha$ be the unique non-negative minimizer.  The
corresponding $\Delta$ in (\ref{dat0}) is also nonnegative, and
satisfies the gap equation (\ref{bcset}).  From this equation it
follows easily that $\Delta$ is strictly positive. In fact, since
$\hat V(p)$ is continuous and strictly negative at the origin, it is
strictly negative in a non-empty open ball around the origin. From
(\ref{bcset}) it follows that $\Delta$ can only vanish at a point
$p_0$ if it vanishes in this open ball centered at $p_0$, and hence is
identically zero.

In particular, from (\ref{dat0}) we conclude that also $\hat\alpha$
is strictly positive. Since we already know that all minimizers must
have the same absolute value, this implies that any (not necessarily
positive) minimizer of $\F_0$ is non-vanishing. But the inequality
(\ref{sq}) is strict for non-vanishing functions, unless
$\hat\alpha(p)=e^{i\kappa} |\hat\alpha(p)|$ for some constant $\kappa
\in \R$. This proves the uniqueness of the minimizer.
\end{proof}

In the following, we shall choose the arbitrary constant phase factor
in the (otherwise) unique minimizer of $\F_0$ as $1$, i.e., we take
$\hat \alpha$ to be positive.

Note that the variational equation for the minimizer of $\F_0$ can be
written as
\begin{equation}\label{evea}
\left(E(-i\nabla) + \lambda V(x) \right)\alpha(x) = 0\,,
\end{equation}
with
$E(p)=|p^2-\mu|/\sqrt{1-4|\hat\alpha(p)|^2}=\sqrt{|p^2-\mu|^2+|\Delta(p)|^2}$,
see (\ref{dat0}) and (\ref{bcset}).  That is, $\alpha$ is an
eigenfunction of the pseudodifferential operator $E(-i\nabla)+\lambda
V(x)$, with zero eigenvalue.  Under our assumptions on $V$, we can
even conclude that it is the ground state.

\begin{Lemma}\label{lem:gs}
  Let $\hat V\leq 0$ and $\hat V(0)<0$. Let $\alpha$ be the minimizer
  of the BCS functional (\ref{deffa}), with corresponding
  $\Delta$ defined in (\ref{dat0}). Then $\alpha$ is the ground state of the operator
  \begin{equation}\label{linop}
    E(-i\nabla) + \lambda V(x)\,,
  \end{equation}
  where $E(p) = \sqrt{(p^2-\mu)^2 + |\Delta(p)|^2}$.
\end{Lemma}

In particular, this implies that $E(-i\nabla)+\lambda V(x) \geq
0$. This property will be an essential ingredient in the proof of
Theorem~\ref{gap}. 

\begin{proof}
  Since $\hat V(p)\leq 0$, the ground state of (\ref{linop}) can be
  chosen to have non-negative Fourier transform. It is therefore not
  orthogonal to $\alpha$, since $\hat\alpha$ is strictly positive by
  Lemma~\ref{uniqueness}. Hence $\alpha$ must be a ground state.
\end{proof}

Similarly to the proof of Theorem~\ref{constant} in
Section~\ref{proof:thm1}, we can employ the Birman-Schwinger principle
to conclude from Lemma~\ref{lem:gs} that $\phi = V^{1/2} \alpha$ satisfies the
eigenvalue equation
\begin{equation}\label{bse}
 \lambda V^{1/2}  \frac1
  {\sqrt{(p^2 - \mu)^2 + |\Delta(p)|^2}}|V|^{1/2} \phi = - \phi\,.
\end{equation}
Moreover, there are no eigenvalues smaller than
$-1$ of the operator on the left side of (\ref{bse}).

Let
\begin{equation}\label{defmtd}
  \widetilde m_\mu(\Delta) = \max\left\{ \frac 1{4\pi\mu} \int_{\R^3}\left( 
\frac 1 {\sqrt{(p^2-\mu)^2 +
        |\Delta(p)|^2}} -\frac 1{p^2}\right) dp \, , \, 0\right\} \,.
\end{equation}
Similarly to (\ref{defmt}), we split the operator in (\ref{bse}) as
$$
V^{1/2} \frac 1{E(-i\nabla)} |V|^{1/2} = \widetilde m_\mu(\Delta) V^{1/2} \dig^*\dig
|V|^{1/2} + V^{1/2}M_\Delta |V|^{1/2}\,.
$$
By proceeding in the same way as in the proof of \cite[Lemma 2]{FHNS},
one shows that $V^{1/2}M_\Delta |V|^{1/2}$ is bounded in the 
Hilbert-Schmidt norm, independently of $\Delta$. Moreover, as in the
proof of Theorem~\ref{constant} (cf.~Eqs.~(\ref{1ba})--(\ref{b1})), the fact that
the lowest eigenvalue of $\lambda V^{1/2} E(-i\nabla)^{-1} |V|^{1/2}$ is $-1$
is, for small enough $\lambda$, equivalent to the fact that the
selfadjoint operator on $L^2(\Omega_\mu)$
\begin{equation}\label{mbd}
 \dig |V|^{1/2} \frac {  \lambda
  \widetilde m_\mu(\Delta)}{1 + \lambda V^{1/2}
    M_\Delta |V|^{1/2} }V^{1/2} \dig^*
\end{equation} 
has $-1$ as its smallest eigenvalue.

Recall that $\dig V \dig^*$ equals $\sqrt\mu\, \V_\mu$ defined in
(\ref{defvm}). Our assumptions on $V$ imply that the lowest eigenvalue
$e_\mu$ of $\V_\mu$ is strictly negative, and non-degenerate. This
implies that $\lim_{\lambda\to 0} \lambda \widetilde m_\mu(\Delta) =
-1/(\sqrt\mu\,e_\mu)$ and hence, in particular, $\widetilde m_\mu(\Delta) \sim
\lambda^{-1}$ as $\lambda \to 0$. The unique eigenfunction
corresponding to the lowest eigenvalue $e_\mu<0$ of $\V_\mu$ is, in
fact, a positive function, and because of radial symmetry of $V$ it is
actually the constant function $u(p)=(4\pi\mu)^{-1/2}$.

We now give a precise characterization of $\Delta(p)$ for small
$\lambda$.

\begin{Lemma}\label{Deltach}
  Let $V \in L^1\cap L^{3/2}$ be radial, with $\hat V \leq 0$ and
  $\hat V(0) < 0$, and let $\Delta$ be given in (\ref{dat0}), with
  $\alpha$ the unique minimizer of the BCS functional
  (\ref{deffa}). Then
  \begin{equation}
    \Delta(p) = - f(\lambda) \left( \int_{\Omega_\mu} \hat V(p-q)  \,
      d\omega(q)  + \lambda \eta_\lambda(p)\right)
  \end{equation}
  for some positive function $f(\lambda)$, with
  $\| \eta_\lambda\|_{L^\infty(\R^3)}$ bounded independently of $\lambda$.
\end{Lemma}

\begin{proof}
  Because of (\ref{bse}), $\dig |V|^{1/2}\phi$ is the
  eigenfunction of (\ref{mbd}) corresponding to the lowest eigenvalue
  $-1$. Note that because of radial symmetry, the constant function
  $u(p)=(4\pi\mu)^{-1/2}$ is an eigenfunction of (\ref{mbd}). For
  small enough $\lambda$ it has to be an eigenfunction corresponding to
  the lowest eigenvalue (since it is the unique ground state of the
  compact operator $\dig V \dig^*$).  We conclude
  that
  \begin{equation}\label{combw}
    \phi =  f(\lambda) \frac 1{1+\lambda V^{1/2} M_\Delta
      |V|^{1/2}} V^{1/2} \dig^* u  = f(\lambda) \left(
      V^{1/2}\dig^* u + \lambda \xi_\lambda\right)
  \end{equation}
  for some normalization constant $f(\lambda)$. Note that
  $\|\xi_\lambda\|_2$ uniformly bounded for small $\lambda$, since
  both $V^{1/2}M_\Delta |V|^{1/2}$ and $V^{1/2}\dig^*$ are bounded
  operators.

Recall from (\ref{dat0}) and (\ref{evea}) and the definition $\phi = V^{1/2}\alpha$ that
$$ 
\Delta(p) = 2 E(p) \hat\alpha(p) = - 2 \lambda \widehat {V
  \alpha}(p) = -2 \lambda \widehat{|V|^{1/2} \phi}(p)\,.
$$
In combination with (\ref{combw}) this implies  that
$$
\Delta(p) = - 2\lambda f(\lambda) \left( \widehat{V \dig^* u}(p) + \lambda
  \widehat{\eta_\lambda}(p)\right)\,,
$$
with $\eta_\lambda = |V|^{1/2} \xi_\lambda$. Note that $\widehat{V
  \dig^* u}(p)$ equals $(4\pi \mu)^{-1/2} \int_{\Omega_\mu} \hat
V(p-q) d\omega(q)$. Moreover, since
$\|\widehat{\eta_\lambda}\|_\infty\leq (2\pi)^{-3/2}
\|\eta_\lambda\|_1 \leq (2\pi)^{-3/2} \|V\|_1 \|\xi_\lambda\|_2$ by
Schwarz's inequality, $\|\widehat {\eta_\lambda}\|_\infty$ is bounded
uniformly in $\lambda$. This implies the statement of the Lemma.
\end{proof}

Note that $ \int_{\Omega_\mu} \hat V(p-q) \, d\omega(q)$ is a
Lipschitz continuous function. In fact,
\begin{align}\nonumber
  \int_{\Omega_\mu} \left(\hat V(p-q)-\hat V(r-q)\right) d\omega(q)
  & = \sqrt \frac{2\mu}\pi \int_{\R^3} V(x) \frac{\sin(\sqrt\mu |x|)}{|x|} \left(
    e^{-ipx}-e^{-irx}\right) dx \\ \nonumber &\leq \sqrt\frac {2\mu}\pi
  |p-r| \int_{\R^3} |V(x)| |\sin(\sqrt{\mu}|x|)| \, dx\,.
\end{align}
Using this property, together with Lemma~\ref{Deltach}, we can now
estimate $\widetilde m_\mu(\Delta)$ in (\ref{defmtd}). We first
consider $\Delta$ to be a constant, and start with the following
observation.

\begin{Lemma}\label{calcmt}
  Let $\vartheta>0$. As $\vartheta\to 0$,
  \begin{equation}
    \widetilde m_\mu(\vartheta)  = \frac 1{4\pi \mu} \int_{\R^3}\left( \frac 1
      {\sqrt{(p^2-\mu)^2 + \vartheta^2}} -\frac 1{p^2}\right) dp  
    = \frac 1{\sqrt\mu}\left( \ln \frac \mu {\vartheta} -2 +
      \ln 8 + o(1)\right) \,.
  \end{equation}
\end{Lemma}

\begin{proof}
  Proceeding as in the proof of Lemma~\ref{calcm}, we have
  \begin{align}\label{split12}
    \widetilde m_\mu(\vartheta) & = \frac 1{2\mu} \int_0^\mu\left(
      \frac{\sqrt{\mu-t} + \sqrt{\mu+t} - 2
        \sqrt{\mu}}{\sqrt{t^2+\vartheta^2}}- \frac{1}{\sqrt{\mu-t}} -
      \frac 1{\sqrt{\mu+t}}\right) dt \\ & \nonumber \quad + \frac
    1\mu \int_0^{\mu} \sqrt{\frac \mu {t^2+\vartheta^2}} dt + \frac
    1{2\mu} \int_\mu^\infty
    \left(\frac{\sqrt{\mu+t}}{\sqrt{t^2+\vartheta^2}} - \frac
      1{\sqrt{\mu+t}}\right) dt \,.
  \end{align}
  The first integral becomes $( \ln 4 -2 -\ln(1+\sqrt{2}))/\sqrt\mu$
  in the limit $\vartheta\to 0$.  The last integral becomes
  $\mu^{-1/2} \ln(1+\sqrt 2)$. Finally,
$$
\int_0^{\mu} \sqrt{\frac \mu {t^2+\vartheta^2}} dt = \sqrt\mu \ln
\frac{\mu + \sqrt{\mu^2 + \vartheta^2}}\vartheta = \sqrt\mu \ln\frac
{2\mu}\vartheta + o(1)\,.
$$
This proves the statement.
\end{proof}

Since $\widetilde m_\mu(\Delta) \to \infty $ as $\lambda\to 0$, if follows
from Lemma~\ref{Deltach} that $\lim_{\lambda\to 0} f(\lambda)=0$. In
the following, we shall slightly abuse the notation and denote by
$\Delta(\sqrt\mu)$ the value of $\Delta(p)$ on the Fermi sphere $\Omega_\mu$. Since
$\Delta$ is a radial function, this is well defined. Using
Lemma~\ref{calcmt}, we will now argue that
\begin{equation}\label{calcmtp}
  \widetilde m_\mu(\Delta) = \frac 1{\sqrt\mu}\left( \ln \frac \mu {\Delta(\sqrt\mu)} -2 +
    \ln 8 + o(1)\right)
\end{equation}
as $\lambda\to 0$. In fact, by inspection of the proof of
Lemma~\ref{calcmt} we see that all we have to show is that
$$
\frac 12 \int_0^{\mu} \frac 1{\sqrt{t^2+\Delta(\sqrt{\mu-t})^2}} dt
+\frac 12 \int_0^{\mu} \frac 1{\sqrt{t^2+\Delta(\sqrt{\mu+t})^2}}
dt = \ln\frac {2\mu}{\Delta(\sqrt\mu)} + o(1)\,,
$$
which follows easily from Lemma~\ref{Deltach}, Lipschitz continuity of
$ \int_{\Omega_\mu} \hat V(p-q) \, d\omega(q)$ and the fact that
$\lim_{\lambda\to 0} f(\lambda)=0$.\

From~(\ref{mbd}) we conclude that
\begin{equation}\label{mbd2}
  \widetilde m_\mu(\Delta)= \frac 1{ \lambda \langle u| \dig V \dig^*|  u\rangle
 - \lambda^2\langle u|  \dig V M_\Delta V  \dig^*| u\rangle
   +O(\lambda^3)}\,,
\end{equation}
where $u(p)=(4\pi\mu)^{-1/2}$ is the normalized constant function on the sphere $\Omega_\mu$. 
Moreover, with $\varphi = V \dig^* u$,
\begin{align}\nonumber
  \langle u| \dig V M_\Delta V \dig^*| u\rangle & = \int_{\R^3} \frac 1{E(p)}
  |\hat\varphi(p)|^2 \, dp - \widetilde m_\mu(\Delta)
  \int_{\Omega_\mu} |\hat\varphi(\sqrt\mu p/|p|)|^2 \, d\omega(p) \\
  \nonumber & = \int_{\R^3} \left(\frac 1{E(p)} \left[ |\hat\varphi(p)|^2
      -|\hat\varphi(\sqrt\mu p/|p|)|^2 \right] + \frac 1{p^2}
    |\hat\varphi(\sqrt\mu p/|p|)|^2 \right) dp \,.
\end{align}
Using Lemma~\ref{Deltach} and the fact that $\lim_{\lambda \to 0}
f(\lambda)=0$, we conclude that
\begin{equation}
 \lim_{\lambda\to 0} \langle u|  \dig V M_\Delta V \dig^* | u\rangle  =  \langle u| \W_\mu| u\rangle \,,
\end{equation}
with $\W_\mu$ defined in (\ref{defW}). (Compare with
Eqs.~(\ref{comeq}) and~(\ref{deno2}).)  In combination with
(\ref{calcmtp}) and (\ref{mbd2}) and the definition of $\B_\mu$ in
(\ref{defB}), this proves that
$$
\lim_{\lambda \to 0} \left(\ln\left(\frac\mu {\Delta(\sqrt\mu)}
  \right) + \frac {\pi}{2 \sqrt{\mu}\, \langle u| \B_\mu|u\rangle }\right) = 2 -
\ln(8)\,.
$$
The same holds true with $\langle u|\B_\mu|u\rangle$ replaced by
$b_\mu(\lambda) = \infspec \B_\mu$, since under our assumptions on $V$
the two quantities differ only by terms of order $\lambda^3$, as
explained in Section~\ref{subsec:ct}.

Now, by the definition of the energy gap $\Xi$ in (\ref{defxi}), $\Xi\leq
\Delta(\sqrt\mu)$. Moreover,
$$
\Xi \geq \min_{|p^2-\mu|\leq \Xi} |\Delta(p)|\,,
$$
from which it easily follows that $\Xi \geq \Delta(\sqrt\mu)( 1-
o(1))$, using Lemma~\ref{Deltach}. This proves Theorem~\ref{gap}.

\section{Conclusion}
We have presented a rigorous analysis of the critical temperature
$T_c$ and the energy gap $\Xi$ for the BCS gap equation. Our results
are valid for a general class of local interaction potentials, and for
any value of the chemical potential. The correct expressions for $T_c$ and
$\Xi$, valid to second order in the Born approximation, do not seem
to have appeared in the literature before.

\section*{Acknowledgments}

The authors would like to thank Pierbiagio Pieri for valuable
discussions. R.S is grateful to the Institute Henri Poincare -- Centre
Emile Borel in Paris for the hospitality and support during part of
this work. R.S. also acknowledges partial support by U.S. National Science
Foundation grant PHY-0652356 and by an Alfred P. Sloan Fellowship.


\end{document}